\documentclass[preprint,aps,floatfix]{revtex4}

\usepackage{amssymb}
\usepackage{graphicx}
\usepackage{dcolumn}
\usepackage{bm}
\usepackage[dvips]{color}
\newcommand{\jpsi}{J/\psi}
\newcommand{\jpsito}{J/\psi \to}
\newcommand{\Lambdabar}{\bar{\Lambda}}
\newcommand{\nc}{\eta_{c}}
\newcommand{\ncto}{\eta_{c} \to}
\newcommand{\Lambdato}{\Lambda \to}
\newcommand{\pim}{\pi^{-}}
\newcommand{\pip}{\pi^{+}}

\begin{document}
\title{Testing the Bell Inequality \\
 at Experiments of High Energy Physics}

\author{Xi-Qing Hao$^1$, Hong-Wei Ke$^2$, Yi-Bing Ding$^3$, Peng-Nian Shen$^4$ and Xue-Qian Li$^1$}
\affiliation{$^1$Department of Physics, Nankai University, Tianjin 300071;\\
$^2$Department of Physics, Tianjin University, Tianjin 300072;\\
$^3$Department of Physics, Graduate University of Chinese Academy of
Sciences, Beijing
100049,\\
$^4$Institute of High Energy Physics, Beijing 100049}

\date{\today} 

\begin{abstract}
Besides using the laser beam, it is very tempting to directly
testify the Bell inequality at high energy experiments where the
spin correlation is exactly what the original Bell inequality
investigates. In this work, we follow the proposal raised in
literature and use the successive decays $J/\psi\to\gamma\eta_c\to
\Lambda\bar\Lambda\to p\pi^-\bar p\pi^+$ to testify the Bell
inequality. Our goal is twofold, namely, we first make a Monte-Carlo
simulation of the processes based on the quantum field theory (QFT).
Since the underlying theory is QFT, it implies that we pre-admit the
validity of quantum picture. Even though the QFT is true, we need to
find how big the database should be, so that we can clearly show
deviations of the correlation from the Bell inequality determined by
the local hidden variable theory. There have been some critiques on
the proposed method, so in the second part, we suggest some
improvements which may help to remedy the ambiguities indicated by
the critiques. It may be realized at an updated facility of high
energy physics, such as BES III.
\end{abstract}

\maketitle

\section{INTRODUCTION \label{sec:intro}}

Since the birth of Quantum Mechanics (QM), dispute about its essence
never ceases. Indeed, nowadays nobody still doubts validity of QM
because of its great success in all fields. However, one can ask if
the theory of QM is complete and the relevant principles, such as
the wave-particle duality, superposition principle and the
probability interpretation are fundamental in nature or just
effective representation of other underlying principles.

Around what is the essence of QM, Bohr and Einstein conducted a
sharp debate over many years. In 1935, Einstein, Podolsky and Rosen
published a collaboration paper where they made a challenge to the
completeness of QM, and it is the famous EPR paradox \cite{EPR}. It
helps to develop the local hidden variable theory (LHVT), especially
Bohm resurrected this field\cite{dbohm}, and then based on this new
understanding, Bell raised a theorem which proves that a local
hidden variable theory cannot repeat all predictions of QM
\cite{Bell}. Because of the local hidden variables, definite
correlations of the involved states would be retained, even though
they are separated by a space-like distance. By contrast, the
quantum entanglement is supposed to be a fundamental character of
nature and it manifests the difference of quantum mechanics from
classical physics. To search a testable scheme which can distinguish
between QM and the LHVT, Bell established the Bell inequality for
correlation among spin polarizations of various states\cite{Bell}.
By the Bell inequality, if the correlation is due to a set of local
hidden variables, the inequality must hold, however, one can easily
show that within a certain parameter range the quantum superposition
of different states would spoil this inequality. Explicitly, in
literature, it is suggested that correlation of polarizations of two
spin-1/2 particles in a singlet state of the total spin, i.e.
$|0,0>$, the Bell inequality holds if the mechanism of local hidden
variable theory (LHVT) applies, whereas it may be explicitly
violated as QM is the dominating mechanism. That is an exclusive and
direct test of LHVT. For almost half century, many experimental
schemes have been designed to realize such a
measurement\cite{Tornqvist,photon,ppcollision,exp}. However, it was
noticed that such experiments demand high accuracy and statistics,
so that become very difficult. Just because of the difficulties, so
far even though the importance of the test is obvious, people are
still unable to carry out experiments with high precision to make a
definite conclusion yet. No wonder, the first success was achieved
in optical experiments, where the correlation between photon
polarizations is studied thanks to application of high quality laser
beams and high precision of optical apparatus as well as advanced
techniques. Over a half century, many experiments have been carried
out to testify, among them, the polarization entanglement
experiments of two-photons and multi-photons attract the widest
attention of the physics society \cite{photon}. All photon
experimental data indicate that the Bell inequality and its
extension forms are violated, and the results are fully consistent
with the prediction of QM. The consistency can reach as high as 30
standard deviations. On other aspect, however, as indicated in
literature, the detection efficiency in optical experiments is
rather low, therefore when analyzing the data, one needs to
introduce additional assumptions, so that the requirement of LHVT
cannot be completely satisfied. That is why as generally considered,
so far, the Bell inequality has not undergone serious test yet.

It would be interesting to return to the original formulism where
correlation between polarizations of two spin-1/2 particles is
discussed. However, because of existence of Lorentz force, the spin
polarization of a charged fermion cannot be directly measured even
though a inhomogenous magnetic field is employed in experiment
\cite{Baym} (we will discuss this problem again at the end of this
paper). Some authors, alternatively suggested to directly measure
the polarization correlation at experiments of high energy
physics\cite{Tornqvist}. An ideal candidate is the successive decay
mode of $\eta_c\to \Lambda\bar\Lambda\to p\pi^-+\bar p\pi^+$,
because $\eta_c$ is a $0^{-+}$ meson and thus $\Lambda\bar\Lambda$
would be in the spin $|0,0>$ state. That is an entangled state of
two spin-1/2 fermions. An obvious advantage is that the information
of the spin polarization of $\Lambda(\bar \Lambda)$ can be obtained
via measuring the directions of the emitted $\pi^-(\pi^+)$, which
would leave clear tracks in the detector.

In the original proposal for testing the Bell inequality, there
should be three independent directions and the polarizations along
all the three directions must be measured and the formulation is
\begin{equation}
|E({\bf a},{\bf b})-E({\bf a},{\bf b'}|)|\leqslant 1+E({\bf b},{\bf
b'}),
\end{equation}
where $E({\bf n_1},{\bf n_2})$ is the correlation of two
polarizations of the two particles which exist in an entangled state
and separated by a space-like distance. As suggested in the original
literature, the inequality with three independent directions is
reduced into a form with only one continuous parameter as
\begin{equation}\label{a1}
E(\theta_{ab})\leqslant 1-\theta_{ab}/\pi,
\end{equation}
where $\theta_{ab}$ is the angle between the linear momenta of the
two emitted pions in the respective CM frames of $\Lambda$ and $\bar
\Lambda$. The authors of Ref.\cite{Tornqvist} showed that the QM
result would upset this inequality.  Obviously, it is a very
difficult experiment which demands a precise measurement. Thanks to
the improvements of detection facility and technique,
experimentalists may be able to measure decay widths of
small-probability processes. Moreover, the BES possesses a largest
database of $J/\psi$ which provides a possibility to test the Bell
inequality at high energy experiments.

On another aspect, this scheme has received some critiques. We will
try to remedy those problems in the last section of the paper.

In this work, it is worth noticing, we simulate the successive
processes $J/\psi\to\eta_c+\gamma$ and $\eta_c\to
\Lambda\bar\Lambda\to p\pi^-+\bar p\pi^+$ based on the quantum field
theory and the aim of such a work is to check if the inequality
(\ref{a1}) could be satisfied. A natural question would be raised
that the Bell inequality is to testify which one of the local hidden
variable theory and QM is valid, thus when we employ the quantum
field theory, it implicitly suggests that we have already assumed
quantum mechanics applying, so that it is not a real test of the
Bell inequality and cannot determine if the local hidden variable
theory fails. In fact, in the first part of this work, we are not
intending to eventually make final conclusion about the inequality,
but will determine how big the database must be to guarantee the
precision for drawing a definite conclusion about difference between
the predictions of QM and local hidden-variable theory. Since QM is
based on statistics, namely all possible quantum states have certain
probabilities to occur, we can only obtain the results with a
certain statistics. Of course, if the event number can be infinity,
the QM would predict a smooth curve for the required relation (the
polarization correlation), but definitely it is impossible, thus a
question emerges. Namely, if the QM is right and the quantum field
theory is a valid theory, we need the number of events to find a
clear distinction above the bound set by the Bell inequality with
one, two or three standard deviations. We use the Monte Carlo method
to simulate the successive processes and draw the graphs with error
bars being explicitly marked out. Finally, we find that the
necessary number of $J/\psi$ must be as large as $10^9$ and when the
statistical and systematic errors of concrete experiments are taken
into account, the database must be enlarged at least by a factor of
10.

Considering the critiques to this method, we propose an improvement
scheme. A long time ago, the authors of Ref.\cite{ppcollision}
suggested to set two Stern-Gerlach apparatuses which provide
inhomogenous magnetic fields to carry out an experiment with two
proton beams to test the Bell inequality. In this work, following
their suggestions, we would set such apparatuses which can
distinguish different spin polarizations of $\Lambda(\bar \Lambda)$,
namely, due to their anomalous magnetic moments, they undergo
different forces and decline their original trajectories. Our
sensitive detector can record their decay daughters' trajectories to
determine their polarizations. Because we have freedom to adjust the
directions of the magnetic fields, we can have three real
independent directions as the original Bell inequality requires. A
detailed discussion will be given in the last section.

This paper is organized as follows. After this long introduction, we
will present all formulae which are necessary for numerical
computations and for readers' convenience, we include a subsection
to discuss the Monte-Carlo simulation and $\chi^2$ analysis. The
third section is for our numerical results where  the Monte-Carlo
errors are given.  In section IV, we discuss our proposal which
suggests a modified version for probing the Bell inequality, and its
advantages and difficulties.

\section{The Bell inequality}

As discussed in the introduction, one of the focuses of basic
research of modern physics is to directly testify the Bell
inequality which is derived based on the LHVT of quantum mechanics,
and it is crucial to justify if QM is a complete local
theory\cite{ding1}.  Starting from the LHVT of quantum mechanics,
one can derive an inequality about the correlation between
polarizations of two spin-1/2 fermions in a system of total-spin
singlet, i.e. the Bell inequality. In the LHVT with the hidden
variable being $\lambda$, the Bell inequality can be phased as
\begin{equation}\label{deq1}
E({\mathbf a},{\mathbf b})=\int d\lambda A({\mathbf
a},\lambda)B({\mathbf b},\lambda)\rho(\lambda),
\end{equation}
where $\rho(\lambda)$ is a distribution of the hidden variable(s)
$\lambda$, $A({\bf a},\lambda)$ is the result of measurement of the
projection (${\bm \sigma^{(A)}}\cdot {\bf a}$) of spin
${\bm\sigma^{(A)}}$  of particle A along direction ${\bf a}$,
whereas $B({\bf b},\lambda)$ is the corresponding measurement for
the projection $({\bm \sigma^{(B)}}\cdot {\bf b})$ of spin ${\bm
\sigma^{(B)}}$ of another particle B along direction ${\bf b}$. The
original expression of Bell inequality refers to three independent
spatial directions ${\bf a}$, ${\bf b}$ and ${\bf c}$ as
\begin{equation}\label{deq2}
    |E({\bf a},{\bf b})-E({\bf a},{\bf c})|-E({\bf b},{\bf
    c})\leqslant 1.
\end{equation}

According to the theory of QM, this correlation function is the
average value of operator (${\bm \sigma^{(A)}}\cdot {\bf a}$)$\cdot$
(${\bm \sigma^{(B)}}\cdot {\bf b})$ over a spin singlet
$|\psi\rangle$ (or written as $|\chi_{00}\rangle$). Namely:
\begin{eqnarray}\label{deq3}
\nonumber\\
E({\bf a},{\bf b})_{\text{QM}} &=& \langle\psi|({\bm
\sigma^{(A)}}\cdot
{\bf a})({\bm \sigma^{(B)}}\cdot {\bf b})|\psi\rangle\nonumber\\
 &=&-{\bf a}\cdot{\bf b}.
\end{eqnarray}
It is easy to prove that the Bell inequality (\ref{deq2})
contradicts to the QM result. If the QM result receives experimental
support, one would quantitatively confirm that the local hidden
variable theory cannot describe all predictions of quantum
mechanics.

Besides the optical experiments, in recent years, some proposals of
non-optical experiments have been raised to testify the LHVT of QM.
Similar tests have also been realized at the proton-proton
double-scattering \cite{ppcollision} and other collider experiments
\cite{exp}, and their conclusions all support the QM and suggest
that the Bell inequality is violated. Comparing with optical
experiments, high energy experiments are dealing with massive
particles, especially the real spin-1/2 fermions. In such processes
weak and strong interactions are involved while only EM interaction
applies in optical experiments.

Among the proposals, we are especially interested in the successive
decay processes $\eta_c\to\Lambda{\bar\Lambda}\to p\pi^- {\bar
p}\pi^+$ at $e^+e^-$ colliders, which was suggested by T\"ornqvist
\cite{Tornqvist} a long while ago. The reason of our interests is
not only because it provides a clear and realizable method, but also
the Beijing Electron-Positron Collider (BEPC) will provide an
incomparably large database which enables people to achieve a high
statistics for drawing a definite conclusion. The spin of $\eta_c$
is zero and decays into a $\Lambda\bar\Lambda$ pair, that is exactly
the Bohm's version about the entangled EPR state, therefore should
serve as an ideal probe for LHVT of QM. In the proposal, the
polarizations of $\Lambda$ and $\bar\Lambda$ can be determined
respectively by the momenta directions of $\pi^+$ and $\pi^-$, and
the relatively longer lifetime of $\Lambda{(\bar\Lambda)}$
guarantees the space-like requirement. All of these set a favorable
condition for directly testing the Bell inequality.

In the standard calculations of the quantum field theory, the
transition matrix element of $\Lambda\to p\pi^-$
reads\cite{Tornqvist}
\begin{equation}\label{deq4}
    M_{\Lambda}=(4\pi)^{1/2}(S+P {\bm \sigma_{\Lambda}}\cdot{\bf a})
\end{equation}
where ${\bf a}$ is the unit vector of $\pi^-$ in the CM frame of
$\Lambda$: ${\bf P_{\pi}^{cm}}\over{|{\bf P_{\pi}^{cm}|}}$, $S$ and
$P$ are the S and P wave-amplitudes respectively. ${\bar\Lambda}\to
{\bar p}\pi^+$ has a similar expression. Thus in the successive
process $\Lambda{\bar\Lambda}\to p\pi^-{\bar p} \pi^+$ the
correlation between the momenta of the two pions can be converted
into the correlation between polarizations of $\Lambda$ and
${\bar\Lambda}$.
\begin{eqnarray}\label{deq5}
\nonumber
    I({\bf a},{\bf b})&=&\langle S|(M_A M_B)^\dag(M_A M_B)|S\rangle\\
    \nonumber\\
    &=&(\frac{|S|^2+|P|^2}{4\pi})^2(1+\alpha^2{\bf a}\cdot{\bf b})
\end{eqnarray}
where $\alpha\approx0.642$.

Except the coefficient $\alpha^2$ and the normalization, this
expression is the same as the correlation function for a system of
two spin-1/2 particles given in the Bohm's version (\ref{deq3}). If
one can neglect the CP violation in the process, the expression can
be understood as: polarization of $\Lambda$ is $\alpha{\bf b}$,
while polarization of ${\bar\Lambda}$ is $\alpha{\bf a}$. Taking
appropriate normalization, the correlation function can be written
as
\begin{equation}\label{deq6}
    R({\bf a},{\bf b})=1+\alpha^2 {\bf a}\cdot{\bf b}
\end{equation}
and the corresponding Bell inequality is rephrased as
\begin{equation}\label{deq7}
    |E({\bf a},{\bf b})-E({\bf a},{\bf c})| \leqslant 1+E({\bf b},{\bf
    c}),
\end{equation}
where $E=\frac{1-R}{\alpha^2}$. T\"ornqvist defined a quantity
$\cos\Theta_{ab}\equiv {\bf a}\cdot{\bf b}$, and by it one notices
that the Bell inequality can be expressed as within a continuous
region, variable $\Theta_{ab}$ satisfies an alternative inequality:
\begin{equation}\label{edq8}
    |E(\Theta_{ab})|\leqslant 1-2\Theta_{ab}/\pi
\end{equation}
\begin{figure}[htb]
\begin{center}
\includegraphics[width=9cm]{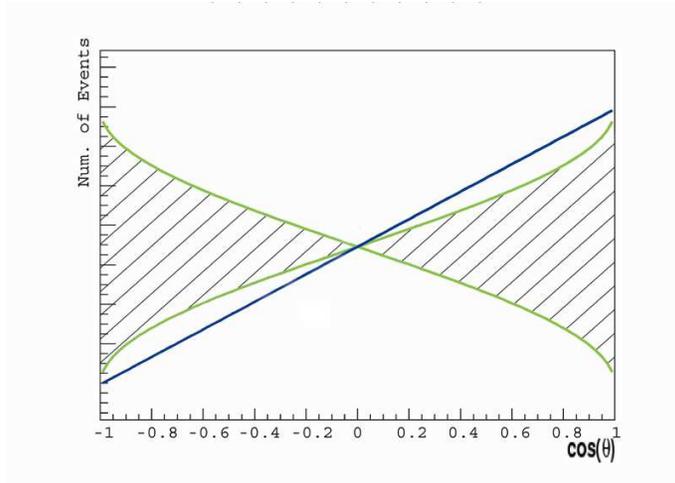}.
\end{center}
\caption{the horizontal axis is $\cos\theta$ and the shadowed region
is allowed by the Bell inequality, and the blue straightline is the
QFT theoretical prediction}. \label{bi}
\end{figure}
Obviously as shown above, by QFT, the correlation function $
|E(\Theta_{ab})|$ is linearly proportional to $\cos\Theta_{ab}$. In
Fig.\ \ref{bi}, for explicitness and clarity, we do not normalize
the data, but keep the event number density where we choose each bin
in the figure covering a region of 0.05($\cos\Theta_{ab}$).
According to T\"ornqvist's suggestion, the DM2 collaboration
reported their preliminary result in 1985. Due to very poor
statistics, they only presented the fitting results for $\jpsi\to
\Lambda\bar\Lambda\to p\pi^-\bar p\pi^+$. Even though in this
process $\Lambda$ and $\bar\Lambda$ also reside in an entangled
state and the correlation functions are similar, there is an obvious
difference from the $\eta_c$ case. Within error tolerance, their
results do not show any obvious discrepancy from the QM prediction,
on other aspect, one still cannot draw any conclusion from the
analysis.

The BES III has begun operation and a great amount of $J/\psi$ and
$\eta_c$ data has been accumulating, so the statistics is
incomparably improved and one is confronting a valuable chance to
make precise measurements on $\eta_c\to \Lambda\bar\Lambda\to
p\pi^-\bar p\pi^+$, just based on this possibility, we will re-carry
out a test on the LHVT at charm energy range. We are going to
simulate the successive decay processes
$J/\psi\to\gamma\eta_c$,$\eta_c\to \Lambda\bar\Lambda\to p\pi^-\bar
p\pi^+$ in terms of the Monte-Carlo method and estimate the
necessary event database for clearly distinguishing QM result from
that obeys the Bell inequality. Even though for a real experimental
setting, this simulation is not accurate enough, it still can
provide a theoretical reference which may guide the experimentalists
to design the experiment, or make decision if one can carry out the
measurement with the considered luminosity and detection error
tolerance at BES III.

\section{Our Method}

\subsection{Random Sampling }

Generally, a complete Monte-Carlo simulation needs, 1. constructing
the concerned random process, 2. taking random numbers which
coincide with the probability function, 3. estimating the errors and
determining the confidence level of the result.

\subsection{Goodness-of-fit Tests of $\chi^2$}

In this scheme, the quantity to be measured is the cosine of the
angle between the two emitted pions. The statistical quantity
\begin{equation}\label{chisquare}
\chi_{obs}^2=\sum\limits_{i=1}^{N}\frac{(n_i-n_{theory})^2}{n_i}=\sum\limits_{i=1}^{N}\frac{(n_i-n
p_{0i})^2}{n_i},
\end{equation}
should approximately obey the $\chi^2(N-1)$ distribution. In the
expression $N$  is the total number of the bins within the concerned
physical region, $n_i$ is the experimental or M-C estimated event
number in the $i$-th bin, $n$ is the total event numbers
(normalization: $\sum n_i=n$), $n p_{0i}$ is the theoretical event
number and $p_{0i}$ is the probability of event appearing in the
i-th bin.

For a given Confidence Level (C.L.), $1 - \alpha$, one can apply the
$\chi^2(N-1)$ accumulation integration to determine the critical
value of $\chi_{\alpha}^2(N-1)$:
\begin{equation}
\alpha = \int_{\chi_{\alpha}^2(N-1)}^{\infty}f(y; N-1)d y.
\end{equation}

In above expression, $f(y; N-1)$ is  the probability density of
$\chi^2(N-1)$ and the critical range is
$\chi^2(N-1)>\chi_{\alpha}^2(N-1)$. Generally, if the theoretically
evaluated value $\chi_{obs}^2$ is greater than
$\chi_{\alpha}^2(N-1)$, it indicates that data are not consistent
with theory at the confidence level $1 - \alpha$; whereas if the
obtained $\chi_{obs}^2$ is smaller than $\chi_{\alpha}^2(N-1)$, at
$1 - \alpha$上 C.L, the data can be described by the theory to be
tested.

\subsection{Simulation and results\label{sec:method}}

We are simulating the successive reactions $\jpsi\to\gamma\eta_c$,
$\eta_c\to\Lambda\Lambdabar$, $\Lambda\to p\pi^-$, $\Lambdabar\to
\bar{p}\pi^+$ step by step. Then taking into account of the Lorentz
boost effects, we calculate the angle between the momenta of the two
emitted pions which are measured in the center-of-mass reference
frames of $\Lambda$ and $\bar\Lambda$ respectively.

In Table.\ \ref{br}, one can note that $BR(\jpsi\to \nc
\gamma)\times BR(\ncto \Lambda \Lambdabar)\times BR(\Lambdato p
\pim)^{2}=5.5\times10^{-6}$, namely, for almost every $10^{6}$
$\jpsi$ events, there would be $5.5$ events corresponding to the
expected reaction $\jpsi \to \nc \gamma\to\Lambda
\Lambdabar\gamma\to p \pim\bar{p} \pip\gamma$. Therefore we first
carry out our M-C simulation with $10^{7}$ $\jpsi$ events, and then
gradually increase the number of the produced $\jpsi$ events.
According to the statistics, we would determine the necessary
$\jpsi$ event number by which we can clearly distinguish the QFT
result from the Bell inequality. In the following figures, the
proportional errors are the reciprocal of the square-root of the
event number in the corresponding bin, i.e. one standard deviation.

\begin{table}[htbp]
\caption{BR of Decay\cite{data}}
\begin{center}
\begin{tabular}{c|c}
\hline
channel & BR　\\
\hline
 $\jpsito \nc \gamma$ & $(1.3\pm0.4)\%$ \\
\hline
$\ncto \Lambda \Lambdabar$  & $(1.04\pm0.31)\times10^{-3}$ \\
\hline
$\Lambdato p \pim$   & $(63.9\pm0.5)\%$  \\
\hline
\end{tabular}
\end{center}
\label{br}
\end{table}

\begin{figure}[htb]
\begin{center}
\includegraphics[width=8cm]{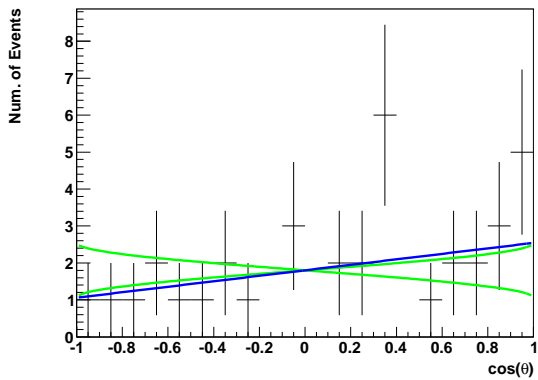}
\end{center}
\caption{ $10^{7}$ $\jpsi$ events, $\theta$ is the angle between
$\pim$ and $\pip$ which are measured respectively in the
center-of-mass reference frames of $\Lambda$ and $\bar\Lambda$. We
divide the region of $\cos\theta$ [-1, +1] into 20 bins. }
\label{bi7}
\end{figure}

\begin{figure}[htb]
\begin{center}
\includegraphics[width=8cm]{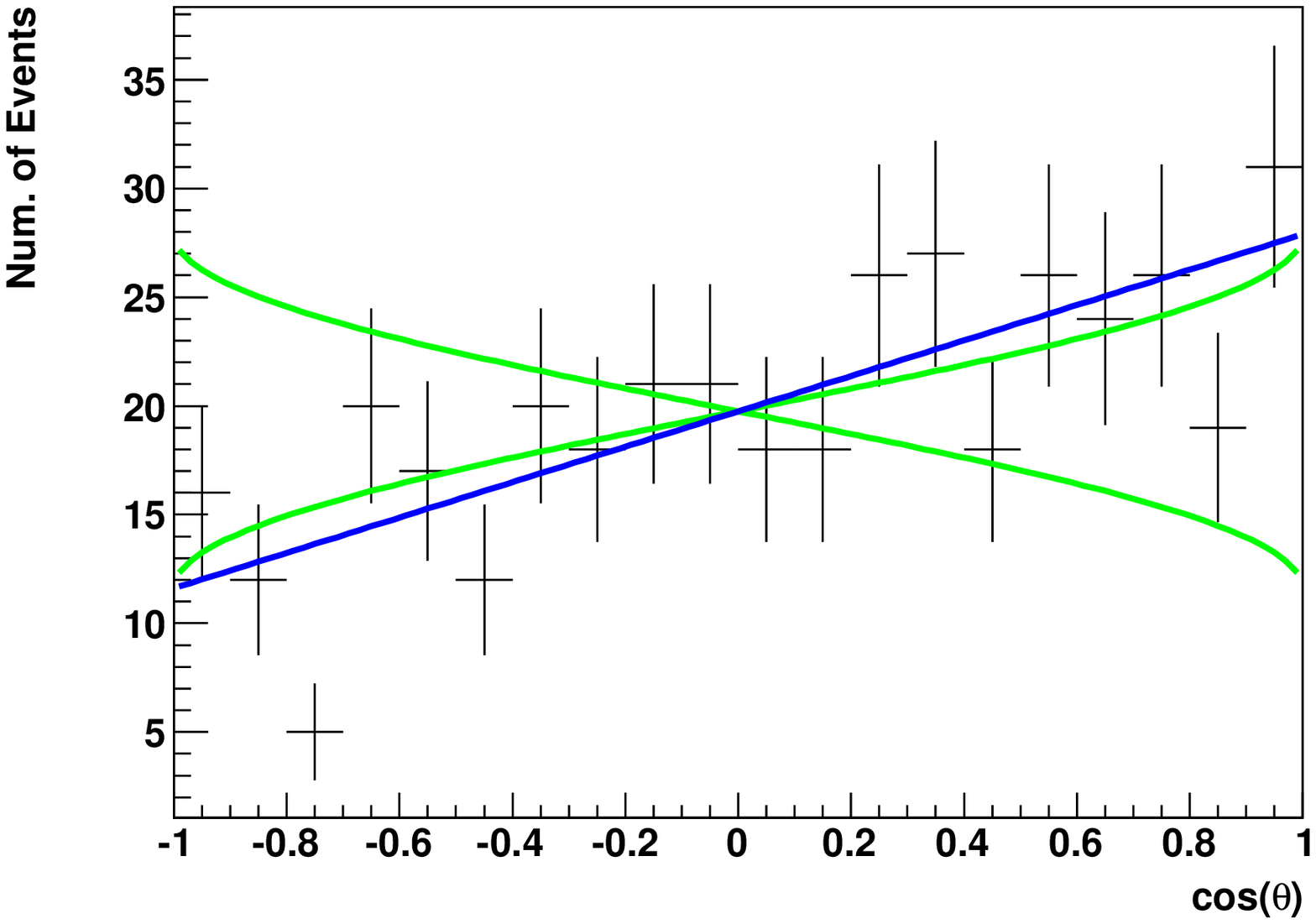}
\end{center}
\caption{ $10^{8}$ $\jpsi$ events, the other notations are the same
as for Fig.\ref{bi7}.} \label{bi8}
\end{figure}

\begin{figure}[htb]
\begin{center}
\includegraphics[width=8cm]{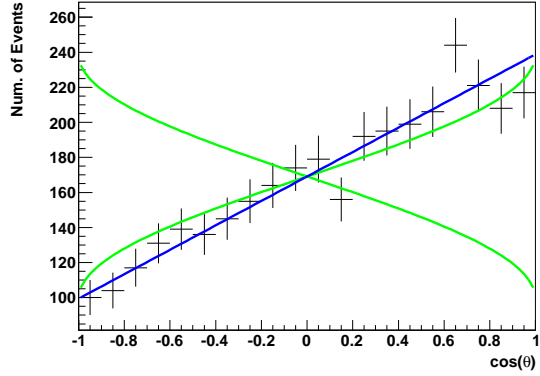}
\end{center}
\caption{ $10^{9}$ $\jpsi$ events, the other notations are the same
as for Fig.\ref{bi7}. } \label{bi9}
\end{figure}

The results of the  Monte-Carlo simulation are shown in Figs.\
\ref{bi7},\ \ref{bi8},\ \ref{bi9}. Then we employ the $\chi^2$
analysis to make the goodness-of-fit test to the boundary values of
the QM prediction in comparison with the Bell inequality. The
corresponding $\chi^2$ values are listed in Table \ref{chisquare}.
The employed equation is $\chi_{Q.M.}^2 =
\sum\limits_{i=1}^{n}(n_i-n_{Q.M.})^2/n_i$, $\chi_{B.I.}^2 =
\sum\limits_{i=1}^{n}(n_i-n_{B.I.})^2/n_i$.

In Fig.\ \ref{bi7}, one can observe that as $10^7$ $\jpsi$ events
are generated, there would only be 36 pion pairs, thus we can note a
left-right asymmetry, but the Monte-Carlo results are not sufficient
to test if QM are different from the Bell inequality.  When the
event number is increased to $10^8$, we obtain $395$ pion pairs, and
the results obviously incline into the prediction of QM (see Fig.\
\ref{bi8}). Even though the errorbars are still too large to clearly
distinguish between the Bell inequality and the QM, by the $\chi^2$
best-fitting test, it is shown that $\chi^2_{Q.M.}<\chi^2_{B.I.}$
and the $\chi^2$ of boundary value of the Bell inequality is larger
than the QM value by 20\%. When the $\jpsi$ event number reaches
$10^9$, we can generate $3382$ pion pairs, the central values of the
Monte-Carlo simulation distribute at the vicinity of the
straightline predicted by the QM, Fig. \ref{bi9}, and the errorbars
also are shortened to the length which is comparable with the
deviation between the QM and Bell inequality. Then the
goodness-of-fit is $\chi^2_{B.I.}/\chi^2_{Q.M.}=1.83$, i.e. the two
theories result in remarkable distinction. It corresponds to 80\%
C.L. Numerical results are shown in the following table. For higher
statistics (smaller statistical errors, oe higher C.L.) much more
$\jpsi$ events are needed.

\begin{table}[htbp]
\caption{$\chi^2$ value of fit with different statistics}
\begin{center}
\begin{tabular}{|c|c|c|c|c|c|}
\hline
statistics ($\jpsi$) & $10^7$& $10^8$ & $10^9$  \\
\hline
$\chi^2_{Q.M.}$ & $7.20$& $27.44$ & $13.07$ \\
\hline
$\chi^2_{B.I.}$ & $7.51$& $32.84$ & $23.89$ \\
\hline
\end{tabular}
\end{center}
\label{chisquare}
\end{table}

As a brief summary, it is possible to test the Bell inequality with
a database of as large as 10$^{9}$ $J/\psi$ events, but the accuracy
is not sufficiently high as expected. Moreover, when detection
efficiency is taken into account, this number must be at least
multiplied by a factor of 10. Thus, for drawing a convincing
conclusion, $10^{11}$ $\jpsi$ events seem to be necessary, and it is
probably beyond the reach of BES III for a few years unless its
luminosity can be enhanced remarkably.

\section{Discussion and our new proposal}

In above section, we carefully investigate possibility of testing
the Bell inequality at $e^+e^-$ collider via successive reactions
$J/\psi\to\gamma\eta_c$,$\eta_c\to \Lambda\bar\Lambda\to p\pi^-\bar
p\pi^+$. With our numerical simulation, we conclude that as the
event number of $\jpsi$ reaches 10$^{9}$, one can test the validity
of the Bell inequality at 80\% C.L., but a real test needs $10^{11}$
$\jpsi$ events at least.

It is worth noticing that our above work does not make a judgement
about validity of the LHVT, but assuming if T\"ornqvist's proposal
really makes sense, how many $\jpsi$ events we need to realize the
test i.e. to effectively distinguish the QM prediction from the
region allowed by the Bell inequality. The analysis can be a
reference for our BES colleagues.

Now let us turn to discuss a different issue which originates from
critiques to the above experiment. According to those critiques, the
reasonability of testing the LHVT with the successive decays
$J/\psi\to\gamma\eta_c$, $\eta_c\to \Lambda\bar\Lambda\to p\pi^-\bar
p\pi^+$ proposed for by T\"orqvist is questionable. Namely, what the
experiment is supposed to do is not the same as the fundamental idea
of Bohm et al. based on analysis on LHVT.

The critiques are focusing on the key point: Can the aforementioned
collider experiments testify if the Bell inequality holds? The
critiques can be summarized into a few points.

\begin{enumerate}
\item The collider experiments are different from the optical
experiments, namely they are not active measurements, but passive
ones, in other words, we cannot control the momentum direction of
the decay products, the pions, and then their polarizations.

\item
Unlike proton and photon, in the collider experiments, the concerned
particles are unstable and only leave tracks of a few cm, which are
too short to let us make precise measurements on the spin
polarizations of the involved particles.

\item The successive reaction such as $J/\psi\to \gamma
\eta_c$,$\eta_c\to\Lambda {\bar\Lambda}\to p\pi^- {\bar p}\pi^+$, is
very complicated, so that the measurements may not realize a concise
test of the original Bell inequality.

\item There may exist a concept substitution\cite{abel}. In QM, different
spin projections are non-commutative, and cannot be simultaneously
measured. In collider experiments, such non-commutative quantities
are replaced by commutative momentum components. That is different
from the LHVT, namely there does not exist a subject to be measured
which is related to the criterion of the EPR completeness. The
reason is that as $\Lambda$ decays, in QM, it corresponds to a
measurement and the momentum represents the result of measurement on
the spin polarizations. Because the measurement induces a
de-coherence, thus part of information is lost.

\item Finally, there is another logic problem.
The spin-polarization correlation of $\Lambda$ and $\bar\Lambda$ is
converted into the correlation between the pion momenta which are
directly measured at collider as $1+\alpha^2{\bf a}\cdot{\bf b}$
where $\bf a$ and $\bf b$ are the directions of the $\pi'$s in the
center-of-mass frames of their respective parent particles
($\Lambda$ or $\bar\Lambda$)(comparing with eq.(8) where their spin
polarizations are respectively $\alpha {\bf a}$  and $\alpha {\bf
b}$ when the CP violation is ignored). However, in the original Bell
inequality the quantity to be dealt with is ${\bf a}\cdot{\bf b}$
instead, thus the region covered by the Bell inequality is $-1\;
to\; +1$ while the direct measured momentum correlation is $
1-\alpha^2 \; to\; 1+\alpha^2$ which has an overlapping region with
[-1,+1]. If one wants to check the distinction between the QM
results with the Bell inequality, he must extract ${\bf a}\cdot{\bf
b}$ from $1+\alpha^2{\bf a}\cdot{\bf b}$ which is derived from QFT
(i.e. QM), thus he is confronting an embarrassing situation that he
needs to admit validity of QM and uses it to testify QM. Generally,
the result loses persuasiveness.

\end{enumerate}

To overcome the problems, some authors \cite{ppcollision} suggested
to carry out the test in terms of the proton-proton
secondary-scattering experiment. They planed to install a
Stern-Gerlach apparatus to measure the spin-polarizations of proton.
However, as indicated in the textbook of QM \cite{Baym}, because
$\nabla \cdot{\bf B}=0$, the Lorentz force would smear out any
observable effect as a non-uniform magnetic field applies.

Inspired by their ideas, we suggest to modify the method proposed by
T\"ornqvist. Our proposal is following. One can install two
Stern-Gerlach \cite{stern-gerlach} apparatuses at two sides with
flexible angles with respect to according to the electron-positron
beams. The apparatus provides a non-uniform magnetic field which may
decline trajectory of the neutral $\Lambda(\bar\Lambda)$ due to its
non-zero anomalous magnetic moment i.e. the force is proportional to
${d\over {\bf n}}(-{\mbox{\boldmath $\mu$}}\cdot {\bf B})$ where
${\mbox{\boldmath $\mu$}}$ is the anomalous magnetic moment of
$\Lambda$, ${\bf B}$ is a non-uniform external magnetic field and
${d\over {\bf n}}$ is a directional derivative. Because $\Lambda$ is
neutral, the Lorentz force does not apply, therefore one may expect
to use the apparatus to directly measure the polarization of
$\Lambda(\bar\Lambda)$. The declination of trajectory of $\Lambda$
(here we use $\Lambda$ as an illustration) depends on its spin
polarization. But one must first identify the particle flying into
the Stern-Gerlach apparatus, i.e. to make sure it is $\Lambda$ or
$\bar\Lambda$. It can be determined by its decay product, i.e. if
its decay product is $p^+\pi^-$, it is $\Lambda$, otherwise is
$\bar\Lambda$. Here one only needs the decay product to tag the
decaying particle, but does not use it to do kinematic measurements.

The advantages are obvious that one can completely avoid the
problems listed above. First, we turn a passive measurement into an
active one, because we can independently adjust direction of each
Stern-Gerlach apparatus, so that we can obtain spin-correlations
among three independent directions as required by the original Bell
inequality. This is in analog to the optical experiment and make
real test on the Bell inequality.

\begin{figure}[htb]
\begin{center}
\includegraphics[width=8cm]{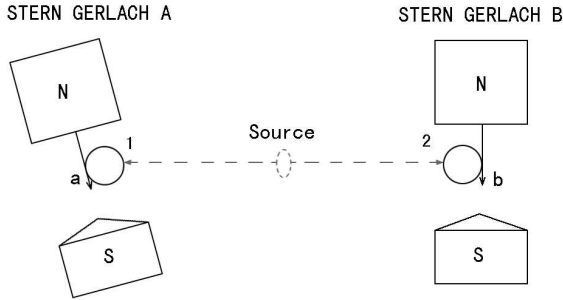}
\end{center}
\caption{ The experimental setting with two Stern-Gerlach
apparatuses } \label{stern}
\end{figure}

One can install such an apparatus as illustrated in Fig.\
\ref{stern}, particles 1 and 2 are produced in a spin singlet and
later are separated by a space-like distance after a time interval
$t$ and then fly into two Stern-Gerlach apparatuses with gradients
of magnetic fields being $\mathbf a$ and $\mathbf b$ respectively.
By their declinations, one can note their polarizations. It
overcomes the flaws of original experimental setting.

There is no free lunch in the world, i.e. advantages also bring up
more difficulties. Let us list a few problems which we can figure
out at present.

\begin{enumerate}\item
The lifetime of $\Lambda$ is not very long, and as products of
$\eta_c$, it can probably travel a few cm before decays, so it would
be difficult to satisfy the space-like conditions. To enforcing the
space-like condition, we need to install the two Stern-Gerlach
apparatuses far enough.  It means that we only choose the
$\Lambda'$s which live longer, if so, the statistics would be
greatly decreased.

\item The magnetic field is not easily adjusted.

\item
The decay products of $\Lambda$ are charged and they may leave very
complicated trajectories in the non-uniform magnetic field and makes
the event reconstruction very difficult. In other words, it is hard
to determine the location of the production vertex.

\item Besides the above difficulties, one still confronts another
serious problem. The magnetic fields in the detector and especially
in the Stern-Gerlach apparatuses may flip the polarizations of
$\Lambda$ and/or $\bar\Lambda$, thus the coherence between $\Lambda$
and $\bar\Lambda$ is spoiled\cite{wang}. Therefore the magnetic
fields in the detector should not be too strong to cause such
decoherence, but on other aspect they cannot be too weak, otherwise
the declinations of the trajectories of $\Lambda$ and $\bar\Lambda$
are not detectable.

\end{enumerate}

Even though all the problems are difficult, we believe that with
rapid developments of high energy physics facilities and detection
techniques they can be eventually solved and we will be able to
testify the LHVT according to its original proposal.\\

\acknowledgments This work was supported in part by the National
Natural Science Foundation of China and the Special Grant for Ph.D
programs of the Education Ministry of China. We are grateful for the
helpful discussions with the members of the Bell Inequality group in
China. One of us (Hao) would like to thank Dr. Y.~L.~Zhu for his
directions on the Monte-Carlo simulations.

%
\end{document}